\documentclass[reqno,12pt,draft]{article}
\usepackage{amssymb,euscript}
\usepackage{amsmath,amssymb}

\newcommand{\e}{\begin{equation}}
\newcommand{\ee}{\end{equation}}
\newcommand{\ea}{\begin{eqnarray}}
\newcommand{\eea}{\end{eqnarray}}

\begin{document}

\begin{flushright}
\end{flushright}
\begin{flushright}
\end{flushright}
\begin{center}

{\Large \textbf{Symplectic  Three-Algebra Unifying ${\cal N}=5,6$
Superconformal Chern-Simons-Matter Theories} \\ }

\bigskip
\textbf{ Fa-Min Chen}\\
{Department of Physics and Astronomy, University of Utah\\
Salt Lake City, UT 84112-0830, USA }
\\
\bigskip
\end{center}

\bigskip
\begin{center}
{\bf Abstract}
\end{center}

We define a 3-algebra with structure constants being
\emph{symmetric} in the first two indices. We also introduce an
invariant anti-symmetric tensor into this 3-algebra and call it a
{\em symplectic } 3-algebra. The general ${\cal N}=5$ superconformal
Chern-Simons-matter (CSM) theory with $SO(5)$ R-symmetry in three
dimensions is constructed by using this algebraic structure. We
demonstrate that the supersymmetry can be enhanced to ${\cal N}=6$
if the sympelctic  3-algebra and the fields are decomposed in a
proper fashion. By specifying the 3-brackets, some presently known
${\cal N}=5, 6$ superconformal theories are described in terms of
this unified 3-algebraic framework. These include the ${\cal N}=5,
Sp(2N)\times O(M)$ CSM theory with $SO(5)$ R-symmetry , the ${\cal
N}=6$, $Sp(2N)\times U(1)$ CSM theory with $SU(4)$ R-symmetry, as
well as the ABJM theory as a special case of $U(M)\times U(N)$
theory with $SU(4)$ R-symmetry.

\newpage

\section{Introduction} \label{Introduction}

Recently, Chern-Simons-matter (CSM) theories with extended
supersymmetries in three dimensions have attracted a lot of
interests, because they are natural candidates of the dual gauge
descriptions of M2 branes. About 20 years ago, generic Chern-Simons
gauge theories (with or without matter) in 3D were demonstrated to
be conformally invariant at the quantum level \cite{CSW1},
\cite{CSW0}, \cite{CSW2}, \cite{Piguet}, \cite{Saemann1}. However,
to describe M2 branes, one needs to further introduce (extended)
supersymmetries into the CSM theories.

The ${\cal N}=8$ CSM theory in $D=3$ with $SO(4)$ gauge group was
first constructed independently by Bagger and Lambert \cite{Bagger}
and by Gustavsson \cite{Gustavsson} (BLG), in terms of the totally
anti-symmetric Nambu 3-brackets \cite{Nambu,
Limiao:1999fm,Bonelli2}. The BLG model is known to be the dual gauge
description of two M2 branes \cite{DMPV, LambertTong, KM:May08}. The
Nambu 3-algebra equipped with a symmetric and positive-definite
metric is essentially unique \cite{Gauntlett, Papadopoulos}: It
generates only an $SO(4)$ gauge symmetry. If, in place of the
symmetric and positive-definite metric, one introduce a Lorentzian
metric, then the so-called Lorentzian 3-algebra can be used to
generate an arbitrary gauge group \cite{Lorentzian3Alg}. However, it
was shown that the BLG model constructed from a Lorentzian
3-algebras is actually an ${\cal N}=8$ super Yang-Mills theory
\cite{GhostFree, CalN8SYM}, not a CSM theory.

A little later, Aharony, Bergman, Jafferis and Maldacena (ABJM) have
been able to construct an ${\cal N}=6$ superconformal CSM theory
with gauge group $U(N)\times U(N)$ and $SU(4)$ R-symmetry
\cite{ABJM}. In their construction, the Nambu 3-brackets did not
play any role. At level $k$, it has been argued that the ABJM theory
describes the low energy limit of $N$ M2-branes probing a
$\textbf{C}^4/\textbf{Z}_k$ singularity. As $k=1, 2$, the
supersymmetry is enhanced to ${\cal N}=8$ \cite{klebanov:Jun09,
Gustavsson:Jun09, KOS:Jun09}. In large-$N$ limit, the ABJM theory
becomes the dual gauge theory of M theory on $AdS_4\times
S^7/\textbf{Z}_k$ \cite{ABJM}. Some further investigation of the
ABJM theory can be found in Ref .\cite{Benna, Schwarz}. In Ref
.\cite{Bergshoeff:2008cz, Bergshoeff}, it has been argued that one
can also obtain the superconformal gauge theories with more or less
supersymmetries by taking a conformal limit of $D=3$ gauged
supergravity theories. Using super Lie algebras to classify the
gauge groups, Gaiotto and Witten (GW) have been able to construct a
large class of ${\cal N}=4$ CSM theories \cite{GaWi}. The GW
theories are extended to include additional twisted hyper-multiplets
\cite{HosomichiJD, Hosomichi:2008jb}. By generalizing Gaiotto and
Witten's construction, two new theories, ${\cal N}=5, Sp(2M)\times
O(N)$ and ${\cal N}=6, Sp(2M)\times O(2)$ CSM theories, were
constructed, and the ABJM theory was re-derived as a special case of
$U(M)\times U(N)$ CSM theories \cite{Hosomichi:2008jb}. The M theory
and string theory dualities of ${\cal N}=5, Sp(2M)\times O(N)$ and
${\cal N}=6,U(M)\times U(N)$ were studied in Ref.
\cite{Aharony:2008gk}.

In an interesting paper, Bagger and Lambert (BL) have been able to
construct the ${\cal N}=6$ ABJM theory in a modified 3-algebra
approach, in which the structure constants are antisymmetric only in
the first two indices \cite{Bagger08:3Alg}. By introducing an
anti-symmetric tensor into a 3-algebra (symplectic 3-algebra), the
authors have constructed another class of ${\cal N}=6$ CSM theories:
the ones with gauge group $Sp(2M)\times O(2)$ \cite{ChenWu1}.
Encouraged by the successes, it is natural to ask whether ${\cal
N}=5$ CSM theories can be constructed in terms of a 3-algebra or
not. Furthermore, it is also natural to ask whether all ${\cal N}=5,
6$ CSM theories can be constructed by a unified 3-algebraic
framework. In this paper we will propose to solve these two
problems.

In section \ref{3AlgN5}, we define a symplectic 3-algebra in which
the structure constants of the 3-brackets are \emph{symmetric} in
the first two indices. The general ${\cal N}=5$ superconformal
Chern-Simons-matter (CSM) theory with $SO(5)$ R-symmetry in three
dimensions is constructed in terms of this symplectic 3-algebra. In
section \ref{Sp2NOM}, we provide the ${\cal N}=5, Sp(2N)\times O(M)$
CSM theory as an example by specifying the 3-brackets. In section
\ref{3AlgN6}, we demonstrate that the supersymmetry can be enhanced
to ${\cal N}=6$ by decomposing the sympelctic 3-algebra and the
fields properly, \emph{and the FI and the symmetry and reality
properties of the structure constants of the ${\cal N}=6$ 3-algebra
can be derived from their ${\cal N}=5$ counterparts}. Therefore all
${\cal N}=5, 6$ superconformal CSM theories are described by a
unified (sympletic) 3-algebraic framework. By specifying the
3-brackets, the ${\cal N}=6$, $Sp(2N)\times U(1)$ and $U(M)\times
U(N)$ CSM are derived in section \ref{Sp2NU1} and \ref{UMUN},
respectively. Especially, the famous ABJM theory is obtained as a
special case of ${\cal N}=6$, $U(M)\times U(N)$ theory.

\textbf{Note added}: Very recently when we were working on the final
version of our manuscript, a work \cite{MFM:Aug09} appeared, which
contains some results overlapping partially with this paper.

\section{Symplectic  3-algebras}
\label{Super3Alg}

It is known that one can construct $D=3, {\cal N}=6$ CSM theories by
using a 3-algebra, in which the structure constants of the
three-bracket are \emph{antisymmetric} only in the first two indices
\cite{Bagger08:3Alg}. This is a natural generalization of the Nambu
3-algebra whose structure constants are totally
antisymmetric\cite{Nambu}. A further generalization would be to
introduce a 3-algebra in which the structure constants are
\emph{symmetric} in the first two indices. In this section, we will
define such a 3-algebra and then use it to construct ${\cal N}=5$
CSM theories in next section.

By definition, a 3-algebra is a complex vector space equipped with a
ternary, trilinear operation, called the 3-bracket. This operation
from three vectors to one vector can be completely determined by its
expressions in terms of a basis (or a set of generators) $T_a$
($a=1,2,\cdots,K$):
\begin{eqnarray}\label{Symp3Bracket}
[T_a,T_b;T_c]=f_{abc }{}^dT_{d },
\end{eqnarray}
where the set of complex numbers $f_{abc }{}^d$ are called the
structure constants. Here we assume that
\begin{equation}\label{symtwo}
[T_a,T_b;T_c]=[T_b,T_a;T_c],
\end{equation}
i.e., the structure constants $f_{abc }{}^d$ are \emph{symmetric} in
the first two indices.

For a field $X$ valued in this 3-algebra, i.e., $X=X^cT_c$, we
define the global transformation of the field as \cite{Bagger}:
\begin{eqnarray}\label{GlbTran}
\delta_{\tilde\Lambda}X=\Lambda^{ab}[T_a,T_b;X],
\end{eqnarray}
where the parameter $\Lambda^{ab}$ is independent of spacetime
coordinate. (We will gauge this symmetry transformation in
subsection \ref{N5CSM}). Because of Eq. (\ref{symtwo}), we require
that $\Lambda^{ab}$ is symmetric in $ab$, i.e.,
$\Lambda^{ab}=\Lambda^{ba}$. Equation (\ref{GlbTran}) is the natural
generalization of $\delta_{\Lambda}X=\Lambda^a[T_a, X]$ in an
ordinary Lie 2-algebra. For an ordinary Lie 2-algebra, the Jacobi
identity is equivalent to
\begin{eqnarray}
\delta_{\Lambda}([X,
Y])=[\delta_{\Lambda}X,Y]+[X,\delta_{\Lambda}Y].
\end{eqnarray}
That is, $\delta_{\Lambda}X=\Lambda^a[T_a, X]$ must act as a
derivative. Analogously, one may require that Eq. (\ref{GlbTran})
acts as a derivative \cite{Bagger}:
\begin{eqnarray}
\delta_{\tilde\Lambda}([X,Y;Z])=
[\delta_{\tilde\Lambda}X,Y;Z]+[X,\delta_{\tilde\Lambda}Y;Z]
+[X,Y;\delta_{\tilde\Lambda}Z]
\end{eqnarray}
Canceling $\Lambda^{ab}, X^e, Y^f$ and $Z^c$ from both sides, the
above equation leads to the following fundamental identity (FI):
\begin{equation}\label{FFI}
[T_a,T_b; [T_e,T_f;T_c]]=[[T_a,T_b;T_e],T_f;
T_c]+[T_e,[T_a,T_b;T_f]; T_c]+[T_e,T_f; [T_a,T_b;T_c]]
\end{equation}
Later we will demonstrate that the FI is equivalent to the
invariance of the structure constants:
$\delta_{\tilde\Lambda}f_{abc}{}^d=0$ (see Eq. (\ref{InvOfStru}))
\cite{Bagger08:3Alg}.

To define a symplectic 3-algebra, we introduce an anti-symmetric
tensor $\omega_{ab}$ and its inverse $\omega^{ab}$ into the
3-algebra. The existence of the inverse of $\omega_{ab}$
($\rm{det}\omega\neq 0$), and the Eq. $\omega_{ab}=-\omega_{ba}$
imply that a 3-algebra index $a$ must run from $1$ to $K=2L$. The
symplectic bilinear form is defined as follows:
\begin{equation}\label{SympInnerProdu}
\omega(X,Y)=\omega_{ab}X^aY^b.
\end{equation}
We require that the above bilinear form to be preserved under
arbitrary global transformations, namely,
\begin{equation}\label{DeltaOmg}
\delta_{\tilde\Lambda}(\omega_{ab}X^aY^b)=0.
\end{equation}
This implies that the structure constants satisfy the condition:
\begin{eqnarray}\label{SymInLast2Ind}
\omega_{de}f_{abc}{}^e=\omega_{ce}f_{abd}{}^e.
\end{eqnarray}
Now the component form of Eq. (\ref{GlbTran}) can be written as
\begin{eqnarray}\nonumber
\delta_{\tilde\Lambda}X^a&=&\Lambda^{bc}f_{bcd}{}^{a}X^d\\ \nonumber
&\equiv&\tilde{\Lambda}^{a}{}_{d}X^d . \label{Transf}
\end{eqnarray}
With the above definition of $\tilde{\Lambda}^{a}{}_{d}$, Eq.
(\ref{DeltaOmg}) must be equivalent to
\begin{eqnarray}\nonumber
\delta_{\tilde\Lambda}\omega_{ab}&=&-\tilde{\Lambda}^c{}_a\omega_{cb}
-\tilde{\Lambda}^c{}_b\omega_{ac}\\ \nonumber
&=&-\Lambda^{de}(f_{dea}{}^c\omega_{cb}+f_{deb}{}^c{}\omega_{ac})\\
&=&0,
\end{eqnarray}
where we used Eq. (\ref{SymInLast2Ind}) in the last line. From point
of view of ordinary Lie group, the (infinitesimal) matrices
$-\tilde{\Lambda}^c{}_a$ are in the Lie algebra of
$Sp(2L,\mathbb{C})$, preserving the anti-symmetric tensor
$\omega_{ab}$ \cite{ChenWu1}. 

By using the FI (\ref{FFI}), one can prove that the structure
constants are also preserved under the global symmetry
transformations \cite{Bagger08:3Alg}:
\begin{eqnarray}\label{InvOfStru}\nonumber
\delta_{\tilde\Lambda}f_{efc}{}^d&=&-\tilde{\Lambda}^g{}_ef_{gfc}{}^d-
\tilde{\Lambda}^g{}_ff_{egc}{}^d-\tilde{\Lambda}^g{}_cf_{efg}{}^d
+\tilde{\Lambda}^d{}_gf_{efc}{}^g\\
 &=&\Lambda^{ab}(-f_{abe}{}^gf_{gfc}{}^d-
f_{abf}{}^g{}f_{egc}{}^d-f_{abc}{}^gf_{efg}{}^d+f_{abg}{}^df_{efc}{}^g)\\
\nonumber &=&0,
\end{eqnarray}
where we have used the FI (\ref{FFI}) in the second line. In other
words, Eq. (\ref{InvOfStru}) is equivalent to the FI (\ref{FFI}).
Thus we can use $\omega_{ab}$ and $f_{abc}{}^d$ to construct
invariant Lagrangians, when the symmetry is gauged.

Later, when we gauge this global symmetry, we require that the gauge fields must be anti-hermitian, leading to a reality condition on the
structure constants (See section (\ref{N5CSM})):
\begin{eqnarray}\label{HermiCondiOnF}
f^*_{abc}{}^d=-\omega^{ae}\omega^{bf}\omega^{cg}\omega_{dh}f_{efg}{}^h.
\end{eqnarray}
Since $\Lambda^{ab}$ carries two symplectic 3-algebra indices, it
obeys the following natural reality condition
\begin{eqnarray}\label{RealCondiOfLmd}
\Lambda^{*ab}=\omega_{ac}\omega_{bd}\Lambda^{cd}.
\end{eqnarray}
Since the 3-algebra is also a complex vector space, there is a
hermitian bilinear form:
\begin{eqnarray}\label{HermiInnerProd}
h(X,Y)=X^{*a}Y^{a}
\end{eqnarray}
(with $X^{*a}$ the complex conjugate of $X^a$) which is
positive-definite and will be used to construct the Lagrangian of
matter fields in CSM theories. The hermitian bilinear form is also
required to be preserved in the sense
\begin{eqnarray}\label{InvOfHermInner}
\delta_{\tilde\Lambda}h(X,Y)=\delta_{\tilde\Lambda}(X^{*a}Y^a)=0.
\end{eqnarray}
As in Ref. \cite{ChenWu1}, we will impose the reality conditions on
the fields valued in the 3-algebra, so that respecting them will
make the anti-symmetric tensor (\ref{SympInnerProdu}) and the
hermitian bilinear form (\ref{HermiInnerProd}) compatible with each
other. Namely the reality conditions essentially require that
 $X^{*a}$ transform in the
same way as $\omega_{ab}X^b$ under the above symmetry
transformations. In fact, by using the reality conditions
(\ref{HermiCondiOnF}) and (\ref{RealCondiOfLmd}), it is easy to
prove that
\begin{eqnarray}\label{covariant}
\delta_{\tilde\Lambda}X^{*a}=\tilde{\Lambda}^{*a}{}_{b}X^{*b}
=-\tilde{\Lambda}^b{}_{a}X^{*b}.
\end{eqnarray}
The last equality indicates that the matrix
$\tilde{\Lambda}^{a}{}_{b}$ is anti-hermitian. Comparing
(\ref{covariant}) with
\begin{equation}
\delta_{\tilde\Lambda}(\omega_{ab}X^b)=-\tilde{\Lambda}^b{}_{a}(\omega_{bc}X^{c}),
\end{equation}
we see that $X^{*a}$ indeed transform in the same way as
$\omega_{ab}X^b$. Therefore, it makes sense to denote $X^{*a}$ as
$\bar{X}_a$, i.e.,
\begin{equation}\label{cmpll}
X^{*a}=\bar{X}_a.
\end{equation}
Also, with (\ref{covariant}), Eq. (\ref{InvOfHermInner}) is
satisfied:
\begin{eqnarray}\label{hmtinv}
\delta_{\tilde\Lambda}(X^{*a}Y^a)&=&(\delta_{\tilde\Lambda}X^{*a})Y^a+X^{*a}(\delta_{\tilde\Lambda}Y^{a})\nonumber\\
&=&-\tilde{\Lambda}^b{}_{a}X^{*b}Y^a+X^{*a}\tilde{\Lambda}^a{}_{b}Y^{b}\nonumber\\
&=&0
\end{eqnarray}
By (\ref{cmpll}), the hermitian bilinear form (\ref{HermiInnerProd})
can be written in a manifest invariant form: \footnote{By our
convention, the hermitian bilinear form is
$h(T_a,T_b)=\delta^a{}_b$. In \cite{Bagger08:3Alg}, it is denoted as
$h_{\bar a b}$, which becomes $\delta_{\bar a b}$ in an orthonormal
basis.}
\begin{equation}
X^{*a}Y^{a}=\bar X_aY^a=\bar X_a\delta^a{}_bY^b,
\end{equation}
and Eq. (\ref{hmtinv}) is equivalent to the following equation:
\begin{eqnarray}\label{InvOfHermInner2}
\delta_{\tilde\Lambda}\delta^a{}_b=\tilde\Lambda^a{}_c\delta^c{}_b-\tilde\Lambda^c{}_b\delta^a{}_c=0.
\end{eqnarray}
In summary, the global transformations (\ref{Transf}) preserve the
hermitian bilinear form (\ref{HermiInnerProd}) and symplectic
bilinear form (\ref{SympInnerProdu}) simultaneously. Or in other
words,
\begin{equation}
\delta_{\tilde\Lambda}\omega_{ab}=0\quad {\rm and}\quad
\delta_{\tilde\Lambda}\delta^a{}_b=0.
\end{equation}
From point of view of ordinary Lie group, the symmetry group
generated by the 3-algebra transformations (\ref{GlbTran}) or
(\ref{Transf}) is the intersection of $U(2L)$ and $Sp(2L,
\mathbb{C})$, which is $Sp(2L)$.

We call the 3-algebra defined by the above Eq. (\ref{Symp3Bracket}),
(\ref{symtwo}), (\ref{FFI}), (\ref{SympInnerProdu}),
(\ref{SymInLast2Ind}), (\ref{HermiInnerProd}) and
(\ref{HermiCondiOnF}) a symplectic 3-algebra. \footnote{We gave the
name `symplectic 3-algebra' in our previous paper \cite{ChenWu1}.}

To construct ${\cal N}=5$ CSM theories, the 3-bracket will be required to
satisfy an additional constraint condition (see section \ref{N5CSM}): \footnote{While we
were writing this paper, the Ref. {\cite{Jose}} appeared which
contains a definition of 3-algebra similar to our definition of
symplectic 3-algebra of this paper. See also \cite{Jose2}. Maybe
there is a connection between our approach and theirs.}
\begin{eqnarray}\label{ConstraintOn3Bracket}
\omega([T_{(a},T_{b};T_{c)}],T_{d})=0
\end{eqnarray}
Or simply $f_{(abc)}{}^e=0$. Now Eq. (\ref{ConstraintOn3Bracket})
implies that $\omega([T_{a},T_{(b};T_{c}],T_{d)})=0$ and
$\omega([T_{a},T_{b};T_{c}],T_{d})=\omega([T_{c},T_{d};T_{a}],T_{b})$.
In summary, the structure constants have the following symmetry
properties:
\begin{eqnarray}\label{SymmeOfF}
\omega_{de}f_{abc}{}^e=\omega_{de}f_{bac}{}^e=\omega_{de}f_{abc}{}^e=\omega_{be}f_{cda}{}^e.
\end{eqnarray}

\section{$D=3,{\cal N}=5$ CSM Theories}
\label{3AlgN5}

\subsection{General ${\cal N}=5$ CSM Theories}\label{N5CSM}

We first postulate that all matter fields are valued in the
symplectic 3-algebra. We then assume the theory has an $SO(5)\cong
Sp(4)$ R-symmetry. It is convenient to use the $Sp(4)$ indices for
$R$-symmetry. We denote the eight complex scalar fields as $Z_A^a$,
and their corresponding complex conjugate $\bar Z_a^A\equiv
Z^{*a}_A$, where $A=1,2,3,4$ transforms in the 4-dimensional
representation of $Sp(4)$, and $a$ is a 3-algebra index. 
Similarly, we
denote the fermion fields and their complex conjugates as
$\psi_A^{a}$ and $\bar\psi_{a}^A$, respectively. The gauge fields
are defined as
\begin{eqnarray}\label{PhysGaugeFld}
\tilde A_\mu{}^c{}_d\equiv A_\mu^{ab}f_{abd}{}^c{},
\end{eqnarray}
where $\mu=0, 1, 2$. Finally, we also impose the reality conditions
on the fields:
\begin{eqnarray}\label{RealCondi}
Z_{A}^{*a}&=&\omega^{AB}\omega_{ab}Z^b_B ,\nonumber\\
\psi_{A}^{*a}&=&\omega^{AB}\omega_{ab}\psi^b_B , \nonumber\\
\tilde A_\mu{}^{*c}{}_d&=&-\omega_{ca}\omega^{db}\tilde
A_\mu{}^a{}_b ,\nonumber\\
A_\mu^{*ab}&=&\omega_{ae}\omega_{bf}A_\mu^{ef}.
\end{eqnarray}
The last two equations of (\ref{RealCondi}) and Eq.
(\ref{PhysGaugeFld}) require that the structure constants obey the
reality condition:
\begin{eqnarray}\label{RealCondiOnF}
f^*_{abc}{}^d=-\omega^{ae}\omega^{bf}\omega^{cg}\omega_{dh}f_{efg}{}^h.
\end{eqnarray}

In terms of the symplectic  3-algebra, now we propose the following
manifestly $Sp(4)$ covariant, ${\cal N}=5$ SUSY transformations:
\begin{eqnarray}\label{GeneSusyTransLaw}\nonumber
\delta Z^a_A&=&i\bar{\epsilon}_A{}^B\psi^a_B\nonumber\\
\delta\psi^a_A&=&\gamma^{\mu}D_\mu Z^a_B\epsilon^B{}_A
+\frac{1}{3}f_{cdb}{}^a\omega^{BC}Z^b_BZ^c_CZ^d_D\epsilon^D{}_A
-\frac{2}{3}f_{cdb}{}^a\omega^{BD}Z^b_CZ^c_DZ^d_A\epsilon^C{}_B
\nonumber\\
\delta \tilde{A}_\mu{}^c{}_d &=&
i\bar{\epsilon}^{AB}\gamma_\mu\psi^b_BZ^a_Af_{abd}{}^c.
\end{eqnarray}
Here $\epsilon^{AB}$ is the antisymmetric supersymmetry parameter,
satisfying
\begin{eqnarray}\label{SusyPara}\nonumber
&&\epsilon^{AB}=-\epsilon^{BA}\nonumber\\
&&\omega_{AB}\epsilon^{AB}=0\nonumber\\
&&\epsilon^{*}_{AB}=\omega^{AC}\omega^{BD}\epsilon_{CD}.
\end{eqnarray}
Namely, they transform as $\bf 5$ of $Sp(4)$. The last equation of
(\ref{SusyPara}) is the reality condition on $\epsilon_{AB}$. The
covariant derivatives are defined as
\begin{eqnarray}
D_\mu Z^A_d &=&
\partial_\mu \bar Z^A_d -\tilde A_\mu{}^c{}_d\bar Z^A_c\\
D_\mu Z_A^d &=&
\partial_\mu Z_A^d +\tilde A_\mu{}^d{}_cZ_A^c.
\end{eqnarray}

Following BL's strategy \cite{Bagger08:3Alg}, we will derive the
equations of motion by requiring that the supersymmetry
transformations are closed on-shell. Let us first examine scalar
supersymmetry transformation. By virtue of the identities in the
appendix \ref{Identities}, we find
\begin{eqnarray}\label{SusyOnZ}
[\delta_{1}, \delta_{2}]Z^{a}_{A}&=&v^{\mu}D_{\mu}Z^{a}_{A}
-\frac{2}{3}f_{bdc}{}^a\Lambda^{cd}Z^{Ab}+\frac{2}{3}
f_{cdb}{}^a\Lambda^{cd}Z^b_A,
\end{eqnarray}
where \begin{eqnarray}v^{\mu} &\equiv& -\frac{i}{2}\bar{\epsilon}_{2}^{BD}\gamma^{\mu}
\epsilon_{1BD},\\
\Lambda^{cd}&\equiv& -\frac{i}{2}Z^{c}_DZ^{d}_{C}
(\bar{\epsilon}_{1}^{CE}\epsilon_{2E}{}^D-\bar{\epsilon}_{2}^{CE}
\epsilon_{1E}{}^D)=\Lambda^{dc},
\end{eqnarray}
and the $\epsilon$ bilinear is symmetric in $CD$. While the first term of
Eq. (\ref{SusyOnZ}) is the gauge covariant translation, we have
to impose some conditions on the structure constants so that the remaining
terms add up to be a gauge transformation. (We will read off the parameter of the gauge transformation by looking the closure of the algebra on the gauge fields.)

We tentatively assume that the third term of Eq. (\ref{SusyOnZ}) is proportional to the gauge transformation. So the second term of Eq. (\ref{SusyOnZ}) should be also proportional to the gauge transformation. This leads us to impose an additional constraint condition on the structure constants:
\begin{eqnarray}\label{ConditionOnF}
\frac{1}{2}(f_{bdc}{}^a+f_{bcd}{}^a)=\frac{\alpha}{2}f_{cdb}{}^a,
\end{eqnarray}
where $\alpha$ is a constant, to be determined later. Now the second and third term of Eq. (\ref{SusyOnZ}) can be combined as
\begin{eqnarray}\label{GaugePrmt1}
\frac{1}{3}(-\alpha+2)f_{cdb}{}^a\Lambda^{cd}Z^b_A,
\end{eqnarray}
which should be the gauge transformation.

Let us now look at the gauge fields:
\begin{eqnarray}
[\delta_1, \delta_2]\tilde{A}_\mu{}^a{}_b&=& v^\nu\tilde{F}_{\nu\mu}{}^a{}_b-(D_\mu\Lambda^{cd})f_{cdb}{}^a \nonumber\\
&&+v^\nu[\tilde{F}_{\mu\nu}{}^a{}_b-\varepsilon_{\mu\nu\lambda}(Z^c_AD^\lambda Z^{Ad}-\frac{i}{2}\bar{\psi}^{Bc}\gamma^\lambda\psi^d_B)f_{cdb}{}^a]\nonumber\\
&&+ {\cal O}(Z^4),\label{SusyOnA}
\end{eqnarray}
where the last term ${\cal O}(Z^4)$ is fourth order in the scalar
fields $Z$. We recognize the second term of the first line as a
gauge transformation \begin{eqnarray}\label{GaugePrmt2}
-(D_\mu\Lambda^{cd})f_{cdb}{}^a=-D_\mu(\Lambda^{cd}f_{cdb}{}^a)
\end{eqnarray}
by a parameter $\tilde{\Lambda}^a{}_b=\Lambda^{cd}f_{cdb}{}^a$,
since the FI (\ref{FFI}) or (\ref{InvOfStru}) implies that $D_\mu
f_{cdb}{}^a=0$ \cite{Bagger08:3Alg}. In accordance with the
parameter, now (\ref{GaugePrmt1}) must satisfy the following
equation: \footnote{According to our convention, if
$\delta_{\tilde{\Lambda}}Z^a_A= \tilde{\Lambda}^a{}_bZ^b_A$, we must
set $\delta_{\tilde{\Lambda}}\tilde{A}_\mu{}^a{}_b=
-D_\mu\tilde{\Lambda}^a{}_b$  so that
$\delta_{\tilde{\Lambda}}(D_\mu Z^a_A)= \tilde{\Lambda}^a{}_b(D_\mu
Z^b_A)$.}
\begin{eqnarray}\label{SolvingAlpha}
\frac{1}{3}(-\alpha+2)f_{cdb}{}^a\Lambda^{cd}Z^b_A=
\Lambda^{cd}f_{cdb}{}^aZ^b_A.
\end{eqnarray}
This equation can be solved by setting $\alpha=-1$. Or in other
words, Eq. (\ref{SolvingAlpha}) can be solved if Eq.
(\ref{ConditionOnF}) can be written as
\begin{eqnarray}\label{RqmtOnF}
f_{(bcd)}{}^a=0,
\end{eqnarray}
which is equivalent to Eq. (\ref{ConstraintOn3Bracket}).
Now Eq. (\ref{SusyOnZ}) becomes
\begin{eqnarray}
[\delta_{1}, \delta_{2}]Z^{a}_{A}&=&v^{\mu}D_{\mu}Z^{a}_{A}+\tilde{\Lambda}^a{}_bZ^b_A,
\end{eqnarray}
as expected.

Following Gustavsson's approach \cite{Gustavsson}, one can
demonstrate that the FI (\ref{FFI}) admits an explicit solution in
terms of a tensor product:
$f_{abc}{}^d=k_{mn}\tau^m_{ab}T^{nd}_{c}$, where $k_{mn}$ is the
Killing-Cartan metric of $Sp(2L)$, and
$\tau^m_{ab}=\omega_{ac}T^{mc}{}_b$ \cite{GaWi}. The matrix
$T^{mc}{}_b$ is in the fundamental representation of $Sp(2L)$, and
$\omega_{ac}$ is the $Sp(2L)$-invariant anti-symmetric tensor. Now
Eq. (\ref{RqmtOnF}) implies that $k_{mn}\tau^m_{(ab}\tau^n_{c)d}=0$,
which is first derived by GW \cite{GaWi}. In the GW theories, it is
the key requirement for enhancing the ${\cal N}=1$ supersymmetry to
the ${\cal N}=4$ supersymmetry.

By using the FI (\ref{FFI}) and the symmetry conditions (\ref{SymmeOfF}), one can prove that the last term of Eq. (\ref{SusyOnA}) vanishes: \begin{eqnarray}
{\cal O}(Z^4)=0.
\end{eqnarray}
So the second line of Eq. (\ref{SusyOnA}) must be the equations of motion for the gauge fields:
\begin{eqnarray}\label{EOMofA}
\tilde{F}_{\mu\nu}{}^a{}_b-\varepsilon_{\mu\nu\lambda}(Z^c_AD^\lambda
Z^{Ad}
-\frac{i}{2}\bar{\psi}^{Bc}\gamma^\lambda\psi^d_B)f_{cdb}{}^a=0.
\end{eqnarray}
Now only the first line of Eq. (\ref{SusyOnA}) remains:
\begin{eqnarray}
[\delta_1, \delta_2]\tilde{A}_\mu{}^a{}_b&=& v^\nu\tilde{F}_{\nu\mu}{}^a{}_b-D_\mu\tilde{\Lambda}^a{}_b,
\end{eqnarray}
which is the desired result.

Finally we turn to the fermion supersymmetry transformation:
\begin{eqnarray}\label{SusyOnPsi}
\nonumber [\delta_1,\delta_2]\psi^a_{A} &=& v^\mu D_\mu
\psi^a_{A} + \tilde{\Lambda}^a{}_{b}
\psi^b_{A}\\
\nonumber &&+\frac{i}{2}(\bar\epsilon_1^{BC}\epsilon_{2BA}
-\bar\epsilon_2^{BC}\epsilon_{1BA})E^a_{C}\\
 &&
-\frac{1}{2}v_\nu\gamma^\nu E^a_{A},
\end{eqnarray}
where
\begin{equation}
E^a_{A} = \gamma^\mu D_\mu\psi^a_{A}
-f_{cdb}{}^aZ^b_BZ^{Bc}\psi^d_A+2f_{cdb}{}^aZ^b_BZ^c_A\psi^{Bd}.
\end{equation}
Hence the equations of motion for fermionic fields are $E^a_{A}=0$. The scalar equations of motion can be derived by taking the super-variation of the fermionic equations of motion:
\begin{eqnarray}
\delta E^a_A=0.
\end{eqnarray}
After Fierz transformation, we obtain two independent parts, containing $\gamma^\mu\epsilon_{BC}$ and $\epsilon_{BC}$, respectively. The part containing $\gamma^\mu\epsilon_{BC}$ merely implies the equations of motion for the gauge fields, so we will not write it down here. The part containing $\epsilon_{BC}$ reads
\begin{eqnarray}\label{EOMofZ}
\bigg(\delta^{[C}_AF^{B]a}+ G_A{}^{BCa}\bigg)\epsilon_{BC}=0,
\end{eqnarray}
where
\begin{eqnarray}
F^{Ba}\equiv
-D^2Z^{Ba}+if_{cdb}{}^aZ^{Cb}\bar{\psi}^{Bc}\psi^d_C+\frac{1}{3}f_{efd}{}^g
f_{gcb}{}^aZ^b_CZ^{Cc}Z^d_DZ^{De}Z^{Bf},
\end{eqnarray}
and
\begin{eqnarray}
\quad G_A{}^{BCa}\epsilon_{BC} &\equiv&
\bigg[if_{cdb}{}^a(\frac{3}{2}Z^{Bd}
\bar{\psi}^{Cc}\psi^b_A+Z^{c}_A\bar{\psi}^{Cb}\psi^{Bd})+if_{d[cb]}{}^aZ^{Bb}
\bar{\psi}^{Cc}\psi^d_A\\
&&+\frac{2}{3}(f_{efd}{}^g f_{gcb}{}^a+f_{ceb}{}^g f_{gfd}{}^a
+2f_{ebd}{}^g
f_{gfc}{}^a)Z^b_DZ^{Dc}Z^{Cd}Z^{Be}Z^f_A\bigg]\epsilon_{BC}.\nonumber
\end{eqnarray}

Since the parameters $\epsilon_{BC}$ are traceless, in the sense that $\omega^{BC}\epsilon_{BC}=\epsilon_{B}{}^B=0$, Eq. (\ref{EOMofZ}) must be equivalent to the following traceless equation:
\begin{eqnarray}
\delta^{[C}_AF^{B]a}-\frac{1}{4}\omega^{BC}F^a_A + G_A{}^{BCa}-\frac{1}{4}\omega^{BC}\omega_{DE}G_A{}^{EDa}=0.
\end{eqnarray}
Contracting on $AC$ gives the scalar equations of motion:
\begin{eqnarray}
F^{Ba}+\frac{4}{5}G_A{}^{BAa}-\frac{1}{5}G^B{}_A{}^{Aa}=0.
\end{eqnarray}
After some simplification we obtain
\begin{eqnarray}
&&0=-D^2 Z^B_a-if_{abc}{}^d(
Z_d^{B}\bar{\psi}^{Cc}\psi^b_C-2Z^{Cc}\bar{\psi}^{b}_C\psi^{B}_d)
\\
&&\quad\quad
-\frac{1}{5}(f_{abc}{}^gf_{gde}{}^f+f_{abd}{}^gf_{gce}{}^f+
3f_{abe}{}^gf_{cdg}{}^f-3f_{abe}{}^gf_{gdc}{}^f)Z^b_AZ^{Ac}Z^d_CZ^{Ce}
Z^{B}_f.\nonumber
\end{eqnarray}

All the equations of motion can be derived as the Euler-Lagrangian equations from the following action:
\begin{eqnarray}\label{GeneN5Lagran}\nonumber
{\cal L}&=&\frac{1}{2}(-D_\mu
\bar Z_a^AD^{\mu}Z^a_A+i\bar{\psi}_a^AD_\mu\gamma^\mu\psi^a_A)\nonumber\\
&&-\frac{i}{2}\omega^{AB}\omega^{CD}\omega_{de}f_{abc}{}^e(Z^a_AZ^c_B\bar{\psi}^b_C\psi^d_D-
2Z^a_AZ^c_D\bar{\psi}^b_C\psi^d_B)\nonumber\\
&&+\frac{1}{2}\epsilon^{\mu\nu\lambda}(\omega_{de}f_{abc}{}^eA_\mu^{ab}\partial_\nu
A_\lambda^{cd}+\frac{2}{3}\omega_{fh}f_{abc}{}^gf_{gde}{}^hA_\mu^{ab}A_\nu^{cd}A_\lambda^{ef})\\
&&-\frac{1}{60}(2f_{abc}{}^gf_{gdf}{}^e-9f_{cda}{}^gf_{gfb}{}^e+2f_{abd}{}^gf_{gcf}{}^e)Z^f_A
Z^{Aa}Z^b_BZ^{Bc}Z^d_C Z^{C}_e.\nonumber
\end{eqnarray}
With the reality conditions (\ref{HermiCondiOnF}) and the first
equation of (\ref{RealCondi}), one can recast the potential term
into the following form:
\begin{equation}
V=\frac{2}{15}(\Upsilon^d_{ABC})^*\Upsilon^d_{ABC},
\end{equation}
where
\begin{equation}
\Upsilon^d_{ABC}\equiv f_{abc}{}^d(Z^a_AZ^b_BZ^c_C
+\frac{1}{4}\omega_{BC}Z^a_AZ^b_DZ^{Dc}).
\end{equation}
Therefore the potential term is actually positive definite. Also it
is not difficult to verify that the Lagrangian (\ref{GeneN5Lagran})
has manifest ${\cal N}=5$ supersymmetry with $Sp(4)$ R-symmetry;
namely it is indeed invariant (up to some boundary terms) under the
supersymmetry transformations (\ref{GeneSusyTransLaw}). It is easy
to check that the above Lagrangian is a scale invariant, local field
theory, provided that the structure constants are dimensionless.
This implies that the theory is classically conformal invariant. We
expect that after quantization it is conformally invariant at the
quantum level.

In the same manner as in our previous paper \cite{ChenWu1}, if we
specify the 3-brackets properly, certain Lie algebra of the gauge
groups can be generated by the FI (\ref{FFI}) of the 3-algebra. In
the next subsection, we will provide the ${\cal N}=5, Sp(2N)\times
O(M)$ CSM theory as an example.

\subsection{${\cal N}=5, Sp(2N)\times O(M)$ CSM
theory}\label{Sp2NOM}

To generate a direct product gauge group, such as $Sp(2N)\times
O(M)$, we first split one 3-algebra index into two indices:
$a\rightarrow k\hat{k}$. As a result, a 3-algebra valued field
becomes $Z^{a}_A \rightarrow Z^{k\hat{k}}_A$. We also decompose the
antisymmetric tensor as $\omega_{ab}\rightarrow
\omega_{\hat{k}\hat{l}}\delta_{kl}$, where $\omega_{\hat{k}\hat{l}}$
is anti-symmetric, and require $Z^{k\hat{k}}_A$ to be valued in the
bi-fundamental representation of $Sp(2N)\times O(M)$. (Here
$k,l=1,\cdots,M$ are the $O(M)$ indices while
$\hat{k},\hat{l}=1,\cdots,2N$ the $Sp(2N)$ indices.) With this
decomposition of $\omega_{ab}$, we can rewrite the reality condition
(\ref{RealCondi}) as
\begin{eqnarray}
Z^{\dag A}_{\hat{k}k}\equiv
\omega^{AB}\omega_{\hat{k}\hat{l}}\delta_{kl}Z^{l\hat{l}}_B,
\end{eqnarray}
and similar conditions for the fermion and gauge fields.
Consequently, the hermitian bilinear form of two fields
\begin{eqnarray}
\omega^{AB}\omega_{ab}Z^b_BZ^a_A= Z^{*a}_AZ^a_A=\bar Z_a^AZ^a_A,
\end{eqnarray}
can be rewritten in a trace form:
\begin{eqnarray}\label{Trace}
Z^{\dag A}_{\hat{k}k}Z^{k\hat{k}}_A= {\rm Tr}(Z^{\dag A}Z_A)
\end{eqnarray}

We then specify the 3-brackets as follows:
\begin{eqnarray}
[T_{k\hat{k}}, T_{l\hat{l}};
T_{m\hat{m}}]=k(\delta_{kl}\omega_{\hat{k}\hat{m}}T_{m\hat{l}}
+\delta_{kl}\omega_{\hat{l}\hat{m}}T_{m\hat{k}}
-\delta_{km}\omega_{\hat{k}\hat{l}}T_{l\hat{m}}
+\delta_{lm}\omega_{\hat{k}\hat{l}}T_{k\hat{m}}).
\end{eqnarray}
The overall coefficient $k$ on the right-hand side is assumed to be
a real constant. It is straightforward to verify that the 3-brackets
satisfy the FI (\ref{FFI}) and the constraints
(\ref{ConstraintOn3Bracket}). The corresponding structure constants
are
\begin{eqnarray}\label{Sp2NOMStru}
f_{\hat{k}k,\hat{l}l,\hat{m}m}{}^{\hat{n}n}
=-k[(\delta_{km}\delta^n_{l}-\delta^n_{k}\delta_{lm})
\omega_{\hat{k}\hat{l}}\delta_{\hat{m}}^{\hat{n}}
-\delta_{kl}\delta_{m}^n(\delta_{\hat{k}\hat{m}}
\delta_{\hat{l}}^{\hat{n}}+\delta_{\hat{k}}^{\hat{n}}
\omega_{\hat{l}\hat{m}})] .
\end{eqnarray}
It is not hard to check that the structure constants have
the symmetry properties (\ref{SymmeOfF}), and satisfy the reality
condition (\ref{RealCondiOnF}). We observe that the structure constants are the same as the components of an embedding tensor in Ref. \cite{Bergshoeff}. This is not merely an accident, and we will explore their relations in a coming paper. With this choice of
structure constants, the gauge fields (\ref{PhysGaugeFld})
become: (We re-scale $A_\mu^{ab}$ by $\frac{1}{k}$ in eq.
(\ref{PhysGaugeFld}).)
\begin{eqnarray}\nonumber
\tilde{A}_{\mu}{}^{m\hat{m}}{}_{n\hat{n}}
&=&A_\mu{}^{k\hat{k},l\hat{l}}
f_{\hat{k}k,\hat{l}l,\hat{n}n}{}^{m\hat{m}}\\
\nonumber &=&-(A_{\mu\hat{nl}}{}^{l\hat{m}}
+A_{\mu}{}^{l\hat{m}}{}_{\hat{n}l})\delta^m{}_n
+(A_{\mu\hat{l}n}{}^{m\hat{l}}
+A_{\mu}{}^{m\hat{l}}{}_{\hat{l}n})\delta^{\hat{m}}{}_{\hat{n}}
\\ \nonumber
&\equiv&-(A_{\mu\hat{n}}{}^{\hat{m}}
+A_{\mu}{}^{\hat{m}}{}_{\hat{n}})\delta^m{}_n +(-B_{\mu
n}{}^m+B_\mu{}^m{}_n)\delta^{\hat{m}}{}_{\hat{n}}\label{PhysGaugeFld1}
\\
&\equiv& \hat{A}_{\mu}{}^{\hat{m}}{}_{\hat{n}}\delta^m{}_n
+A_\mu{}^m{}_n\delta^{\hat{m}}{}_{\hat{n}}.
\end{eqnarray}
It is easy to see that $\hat{A}_{\mu}{}^{\hat{m}}{}_{\hat{n}}$ is
the $Sp(2N)$ part of the gauge potential, because it can be written
as
$A_{\mu}^{\hat{k}\hat{l}}(t_{\hat{k}\hat{l}}){}^{\hat{m}}{}_{\hat{n}}$,
where $(t_{\hat{k}\hat{l}}){}^{\hat{m}}{}_{\hat{n}}$ is the
fundamental representation of the ordinary Lie algebra $Sp(2N)$.
Similarly, we can identify $A_\mu{}^m{}_n$ as the $O(M)$ part of the
gauge potential. As we explained in our previous paper
\cite{ChenWu1}, the Lie algebra of the gauge group $Sp(2N)\times
O(M)$ is actually generated by the FI (\ref{FFI}) after we specify
the structure constants by Eq. (\ref{Sp2NOMStru}).

We would like to derive the ${\cal N}=5, Sp(2N)\times O(M)$ Lagrangian
and the corresponding supersymmetry transformation law in the 3-algebraic framework. With the notation (\ref{Trace}), the kinetic terms for matter fields in the Lagrangian (\ref{GeneN5Lagran}) read
\begin{eqnarray}
-\frac{1}{2}{\rm Tr}(D_\mu Z^{\dag A}D^\mu
Z_A-i\bar{\psi}^{\dag A}D_\mu\gamma^\mu\psi_A ).
\end{eqnarray}

With the choice of the structure constants (\ref{Sp2NOMStru}), we learn that
\begin{equation}
\omega_{de}f_{abc}{}^eX^aY^bZ^cW^d=-k\rm{Tr}(XY^\dag ZW^\dag+YX^\dag
ZW^\dag-ZX^\dag YW^\dag-ZY^\dag XW^\dag).
\end{equation}
Hence the Yukawa terms in the Lagrangian (\ref{GeneN5Lagran}) become
\begin{eqnarray}
&&ik
\varepsilon^{ABCD}{\rm Tr}(Z_A\bar{\psi}^\dag_BZ_C\psi^\dag_D)
\\ \nonumber
&&-i\frac{k}{2}{\rm Tr}(\bar{\psi}^\dag_AZ_BZ^{\dag
B}\psi^A-\bar{\psi}_AZ^{\dag}_BZ^B\psi^{\dag
A}-2\bar{\psi}^\dag_AZ_BZ^{\dag A}\psi^B+2\bar{\psi}^AZ^{\dag
B}Z_A\psi^\dag_B),
\end{eqnarray}
where we have used the following $Sp(4)$ identity:
\begin{eqnarray}\label{Sp4Id1}
\varepsilon^{ABCD}&=&-\omega^{AB}\omega^{CD}
+\omega^{AC}\omega^{BD}-\omega^{AD}\omega^{BC}.
\end{eqnarray}

Substituting the definition of the gauge fields
(\ref{PhysGaugeFld1}) into the `twisted' Chern-Simons term in the
Lagrangian (\ref{GeneN5Lagran}) gives the conventional Chern-Simons
term
\begin{eqnarray}
\frac{1}{4k}\epsilon^{\mu\nu\lambda}{\rm Tr} (\hat
A_\mu\partial_\nu\hat A_\lambda+\frac{2}{3}\hat A_\mu\hat A_\nu\hat
A_\lambda-A_\mu\partial_\nu A_\lambda-\frac{2}{3}A_\mu A_\nu
A_\lambda).
\end{eqnarray}

Finally we want to calculate the potential terms in the Lagrangian
(\ref{GeneN5Lagran}). By using
$\omega_{de}f_{abc}{}^e=\omega_{ce}f_{abd}{}^e$, they can be
re-written as
\begin{eqnarray}\nonumber
&&\frac{1}{60}{\rm Tr}(2[Z^A,Z_B;Z^B][Z^C,Z_A;Z_C]^\dag-9[Z^B,Z_C;Z^A][Z^C,Z_B;Z_A]^\dag
\\&&\quad\quad\quad+2[Z^A,Z_B;Z_C][Z^C,Z_A;Z^B]^\dag).
\end{eqnarray}
The last two terms can be combined together:
\begin{eqnarray}\nonumber
&&-\frac{4k^2}{15}(\omega_{AF}\omega_{BE}\omega_{CD}-2\omega_{AF}\omega_{BC}\omega_{DE}
+2\omega_{AC}\omega_{BF}\omega_{DE}
-\omega_{AE}\omega_{BF}\omega_{CD}\\
&&\quad\quad\quad+\omega_{AC}\omega_{BE}\omega_{DF}-\omega_{AE}\omega_{BC}\omega_{DF}){\rm Tr}(Z^BZ^{\dag D}Z^AZ^{\dag C}Z^EZ^{\dag F}).
\end{eqnarray}
The first term becomes
\begin{eqnarray}
&&\frac{k^2}{30}(2\omega_{AD}\omega_{BE}\omega_{CF}+4\omega_{AB}\omega_{CF}\omega_{DE}
-2\omega_{AE}\omega_{BD}\omega_{CF}+\omega_{AD}\omega_{BC}\omega_{EF}\nonumber\\
&&\quad\quad-2\omega_{AB}\omega_{CD}\omega_{EF}-\omega_{AC}\omega_{BD}\omega_{EF}
+\omega_{AD}\omega_{BF}\omega_{CE}+2\omega_{AB}\omega_{CE}\omega_{DF}\nonumber\\
&&\quad\quad-\omega_{AF}\omega_{BD}\omega_{CE}){\rm Tr}(Z^BZ^{\dag D}Z^AZ^{\dag C}Z^EZ^{\dag F}).
\end{eqnarray}
Clearly, they can be simplified further. Taking account of the cyclic property of the trace, there are only four possible
potential terms:
\begin{eqnarray}
&&(c_1\omega_{AD}\omega_{BE}\omega_{CF}+c_2\omega_{BD}\omega_{CE}\omega_{AF}
+c_3\omega_{AD}\omega_{CE}\omega_{BF}
+c_4\omega_{CD}\omega_{AE}\omega_{BF})\nonumber\\&&\quad\times{\rm Tr}(Z^AZ^{\dag D}Z^BZ^{\dag E}Z^CZ^{\dag F}),
\end{eqnarray}
where $c_1,\cdots$ and $c_4$ are constants. After some work, we reach the final expression for the potential:
\begin{eqnarray}\nonumber
&&\frac{k^2}{6}{\rm Tr}(-6Z_AZ^{\dag A}
Z_BZ^{\dag C}Z_CZ^{\dag B} +4Z_AZ^{\dag
C}Z_BZ^{\dag A}Z_CZ^{\dag B} \\
&&\quad\quad\quad\quad+Z_AZ^{\dag A}Z_BZ^{\dag B}Z_CZ^{\dag C}
+Z_AZ^{\dag B}Z_BZ^{\dag C}Z_CZ^{\dag A}).
\end{eqnarray}
In deriving this potential, we have used another $Sp(4)$ identity
\cite{Hosomichi:2008jb}:
\begin{eqnarray}
\label{Sp4Id2} \varepsilon_{GABC}\varepsilon^{GDEF}
&=&3!\delta_{[A}^D\delta_B^E\delta_{C]}^F\\ \nonumber
&=&3(-\delta_{[A}^D\omega^{EF}\omega_{BC]}
+\delta_{[A}^E\omega^{DF}\omega_{BC]}
-\delta_{[A}^F\omega^{DE}\omega_{BC]}) .
\end{eqnarray}
In summary, with the choice of the structure constants (\ref{Sp2NOMStru}),
the Lagrangian (\ref{GeneN5Lagran}) is given by
\begin{eqnarray}\label{Sp2NOMSLagran}\nonumber
{\cal L}&=&-\frac{1}{2}{\rm Tr}(D_\mu Z^{\dag A}D^\mu
Z_A-i\bar{\psi}^{\dag A}D_\mu\gamma^\mu\psi_A ) +ik
\varepsilon^{ABCD}{\rm Tr}(Z_A\bar{\psi}^\dag_BZ_C\psi^\dag_D)
\\ \nonumber
&&-i\frac{k}{2}{\rm Tr}(\bar{\psi}^\dag_AZ_BZ^{\dag
B}\psi^A-\bar{\psi}_AZ^{\dag}_BZ^B\psi^{\dag
A}-2\bar{\psi}^\dag_AZ_BZ^{\dag A}\psi^B+2\bar{\psi}^AZ^{\dag
B}Z_A\psi^\dag_B)
\\ \nonumber
&&+\frac{1}{4k}\epsilon^{\mu\nu\lambda}{\rm Tr} (\hat
A_\mu\partial_\nu\hat A_\lambda+\frac{2}{3}\hat A_\mu\hat A_\nu\hat
A_\lambda-A_\mu\partial_\nu A_\lambda-\frac{2}{3}A_\mu A_\nu
A_\lambda) \\ \nonumber &&+\frac{k^2}{6}{\rm Tr}(-6Z_AZ^{\dag A}
Z_BZ^{\dag C}Z_CZ^{\dag B} +4Z_AZ^{\dag
C}Z_BZ^{\dag A}Z_CZ^{\dag B} \\
&&\quad\quad\quad\quad+Z_AZ^{\dag A}Z_BZ^{\dag B}Z_CZ^{\dag C}
+Z_AZ^{\dag B}Z_BZ^{\dag C}Z_CZ^{\dag A}),
\end{eqnarray}

Substituting the structure constants $(\ref{Sp2NOMStru})$ into
$(\ref{GeneSusyTransLaw})$, the SUSY transformation law reads
\begin{eqnarray}\label{Sp2NOMSusyTrans}\nonumber
\delta Z_A&=&i\bar{\epsilon}_A{}^B\psi_B\nonumber\\
\delta\psi_A&=&\gamma^{\mu}D_\mu
Z_B\epsilon^B{}_A-\frac{2k}{3}\epsilon^C{}_A(Z_{[B}Z^{\dag
B}Z_{C]}+Z_BZ^\dag_CZ^B)\nonumber\\&&+\frac{4k}{3}\epsilon^C{}_B(Z_{[C}Z^{\dag
B}Z_{A]}+Z_CZ^\dag_AZ^B)\nonumber\\
\delta A_\mu{}&=&ik\bar{\epsilon}^{AB}\gamma_\mu
(Z_A\psi^\dag_B+\psi_BZ^\dag_A)\nonumber\\ \delta
\hat{A}_\mu{}&=&-ik\bar{\epsilon}^{AB}\gamma_\mu
(\psi^\dag_BZ_A+Z^\dag_A\psi_B).
\end{eqnarray}
The ${\cal N}=5, Sp(2N)\times O(M)$ Lagrangian (\ref{Sp2NOMSLagran})
and the supersymmetry transformation law (\ref{Sp2NOMSusyTrans}) are
in agreement with those given in ref. \cite{Hosomichi:2008jb}, which
were derived in terms of ordinary Lie algebra. This theory has been
conjectured to be the dual gauge theory of M2 branes probing a
$\textbf{C}^4/\hat{\textbf{D}}_k$ singularity, where
$\hat{\textbf{D}}_k$ is the binary dihedral group
\cite{Hosomichi:2008jb, Aharony:2008gk}.

\section{$D=3, {\cal N}=6$ CSM Theories from  3-algebras}
\label{3AlgN6}

In Ref. \cite{Hosomichi:2008jb}, the ${\cal N}=6$ theories are
derived from the ${\cal N}=5$ theories by enhancing the R-symmetry
from $Sp(4)$ to $SU(4)$. In this section we will implement the same
idea in the context of 3-algebras. We will call the symplectic
3-algebras presented in this paper and in Ref. \cite{Bagger08:3Alg},
respectively, to construct the ${\cal N}=5$, ${\cal N}=6$ theories
as the ``${\cal N}=5$, ${\cal N}= 6$ 3-algebra", respectively. We
will see that the symplectic 3-algebra provides framework unifying
the ${\cal N}=5$ and ${\cal N}=6$ CSM models.

\subsection{General ${\cal N}$=6 CSM Theories}
\label{N6}

The enhancement of R-symmetry from $Sp(4)$ to $SU(4)$ in ref.
\cite{Hosomichi:2008jb} is based on the following observation: The
reality condition (\ref{RealCondi}) implies that the complex
conjugate of a matter field can be obtained by a similarity
transformation. Therefore the matter fields actually furnish a
pseudo-real presentation of the gauge group. If we decompose this
pseudo-real representation into a complex representation and its
conjugate representation, then the $Sp(4)$ R-symmetry will be
enhanced to $SU(4)$, and the global ${\cal N}=5$ SUSY will get
enhanced to ${\cal N}=6$.

In this section, we will show that this enhancement can be
implemented exclusively in the framework of symplectic 3-algebra,
which thus provides a unified framework for both ${\cal N}=5$
and ${\cal N}=6$ theories. Since in our approach the ordinary Lie
algebra of the gauge groups is generated by the FI and the
3-brackets, the challenge we face is to derive the ${\cal N}=6$
3-algebra from the 3-algebra proposed in this paper.

Following ref. \cite{Hosomichi:2008jb}, we first decompose an ${\cal
N}=5$ scalar field as a direct sum of an ${\cal N}=6$ scalar field
and its complex conjugate (See eq. (\ref{N6RealCondi})):
\begin{eqnarray}\label{DecomScaFld}
(Z_A^a)_{{\cal N}=5}\rightarrow Z_A^{a\alpha}=
\bar{Z}^a_A\chi_{1\alpha}+
\omega_{AB}Z^B_a\chi_{2\alpha}=\bar{Z}^a_A\delta_{1\alpha}+
\omega_{AB}Z^B_a\delta_{2\alpha} ,
\end{eqnarray}
where the right hand side of the arrow contains ${\cal N}=6$ fields.
Here the index $a$ of the left hand side of the arrow runs from $1$
to $2L$, while the index $a$ of the right hand side of the arrow
runs from $1$ to $L$. And $\chi_{1\alpha}$ and $\chi_{2\alpha}$ are
``spin up" and ``spin down" spinor, respectively; i.e.,
\begin{eqnarray}
\chi_{1\alpha}=\begin{pmatrix} 1 \\ 0
\end{pmatrix}\;,
\;\;\;\;\; \chi_{2\alpha}=\begin{pmatrix} 0 \\ 1
\end{pmatrix}.
\end{eqnarray}
To make the ${\cal N}=5$ SUSY transformation law
(\ref{GeneSusyTransLaw}) consistent with that of ${\cal N}=6$ (see
below the first two equations of eq. (\ref{N6susy})), we have to
decompose the ${\cal N}=5$ fermion fields as follows:
\begin{eqnarray}\label{DecomFerFld}
(\psi^a_A)_{{\cal N}=5}\rightarrow
\psi^{a\alpha}_A=\omega_{AB}\psi^{Ba}\delta_{1\alpha}
-\psi_{Aa}\delta_{2\alpha} ,
\end{eqnarray}
where the right hand side contains ${\cal N}=6$ fermion fields. We
further decompose the anti-symmetric tensor $\omega_{ab}$ and its
inverse as
\begin{eqnarray}\label{DecomMetr}\nonumber
\omega_{ab}\rightarrow
\omega_{a\alpha,b\beta}=\delta_a{}^b\delta_{1\alpha}\delta_{2\beta}
-\delta^a{}_b\delta_{2\alpha}\delta_{1\beta},\\
\omega^{ab}\rightarrow \omega^{a\alpha,b\beta}
=\delta_a{}^b\delta_{2\alpha}\delta_{1\beta}
-\delta^a{}_b\delta_{1\alpha}\delta_{2\beta} \ .
\end{eqnarray}
Then the reality condition (\ref{RealCondi}) reads
\begin{eqnarray}\label{N6RealCondi}
Z^{*A}_a= \bar Z_A^{a},\quad\quad \psi^{*Aa}=\psi_{Aa},
\end{eqnarray}
in agreement with those for ${\cal N}=6$ theories. This justifies
the above decomposition (\ref{DecomMetr}) of the anti-symmetric
tensor of the ${\cal N}=5$  3-algebra to derive the ${\cal N}=6$
3-algebra.

To be compatible with the decomposition of scalar and fermion
fields, one has to decompose the gauge fields as
\begin{eqnarray}\label{N6GaugFld}
(\tilde{A}_\mu{}^a{}_b)_{{\cal N}=5}\rightarrow
\tilde{A}_\mu{}^{a\alpha}{}_{b\beta}=
\tilde{A}_\mu{}^a{}_b\delta_{1\alpha}\delta_{1\beta}
-\tilde{A}_\mu{}^b{}_a\delta_{2\alpha}\delta_{2\beta} ,
\end{eqnarray}
where the right hand side is a direct sum of ${\cal N}=6$ gauge
fields and their complex conjugates. Since our gauge fields
$(\tilde{A}_\mu{}^c{}_d)_{{\cal N}=5}$ are defined in terms of the
structure constants of a 3-algebra, i.e.,
\begin{eqnarray}
(\tilde{A}_\mu{}^c{}_d)_{{\cal
N}=5}=(A_\mu^{ab}f_{abd}{}^{c})_{{\cal N}=5} ,
\end{eqnarray}
(see also Eq. (\ref{PhysGaugeFld})), we have to decompose its
structure constants properly to result in the desired decomposition
Eq. (\ref{N6GaugFld}). We find that Eq. (\ref{N6GaugFld}) indeed
follows from the decomposition of the structure constants given by
\begin{eqnarray}\label{DecomStrucConst}\nonumber
(\omega_{de}f_{abc}{}^e)_{{\cal N}=5}&\rightarrow& \omega_{d\delta,e\eta}f_{a\alpha,b\beta,c\gamma}{}^{e\eta}\nonumber\\
&=&f^{ac}{}_{db}\delta_{2\alpha}\delta_{1\beta}\delta_{2\gamma}\delta_{1\delta}
+f^{ad}{}_{cb}\delta_{2\alpha}\delta_{1\beta}\delta_{1\gamma}\delta_{2\delta}\nonumber\\
&&+f^{bc}{}_{da}\delta_{1\alpha}\delta_{2\beta}\delta_{2\gamma}\delta_{1\delta}
+f^{bd}{}_{ca}\delta_{1\alpha}\delta_{2\beta}\delta_{1\gamma}\delta_{2\delta},
\end{eqnarray}
combined with the decomposition of $(A_\mu^{ab})_{{\cal N}=5}$ given
by
\begin{eqnarray}\label{DecomGauFld}
(A_\mu^{ab})_{{\cal N}=5}\rightarrow
A_\mu^{a\alpha,b\beta}=-\frac{1}{2}(A_\mu{}^a{}_b\delta_{1\alpha}\delta_{2\beta}
+A_\mu{}^b{}_a\delta_{2\alpha}\delta_{1\beta}) .
\end{eqnarray}
With these decompositions, the ${\cal N}=6$ gauge fields become:
(see the right side of Eq. (\ref{N6GaugFld}))
\begin{eqnarray}\label{N6PhyGaugFld}
\tilde{A}_\mu{}^{c}{}_{d}=A_\mu{}^b{}_af^{ca}{}_{bd} .
\end{eqnarray}
Later we will identify the above $f^{ca}{}_{bd}$ in the right side
of eq. (\ref{DecomStrucConst}) as the structure constants of the
${\cal N}=6$ 3-algebra. With eq. (\ref{DecomMetr}) and
(\ref{DecomStrucConst}), the reality condition of the structure
constants (\ref{RealCondiOnF}) reduces to
\begin{eqnarray}\label{RealCondiOnN6F}
f^{*ab}{}_{cd}=f^{cd}{}_{ab} ,
\end{eqnarray}
as desired for the ${\cal N}=6$ 3-algebra \cite{Bagger08:3Alg,
ChenWu1}.

Since we decompose a matter field as a direct sum of a ${\cal N}=6$
matter field and its \emph{complex conjugate}, it is necessary to
decompose a generator of the  3-algebra as a direct sum of a
generator of a 3-algebra and its \emph{complex conjugate}. This can
be accomplished by setting
\begin{eqnarray}
(T_{a})_{{\cal N}=5}\rightarrow T_{a\alpha} &=&\bar
t_{a}\delta_{1\alpha}-t^{a}\delta_{2\alpha} ,
\end{eqnarray}
where $t^a$ is a generator of the 3-algebra, and $\bar{t}_a$ its
complex conjugate.

The hermitian bilinear form of two ${\cal N}=5$ fields will be (for
instance):
\begin{eqnarray}\nonumber
(Z^{*a}_{1A}Z_{2A}^a)_{{\cal N}=5}
&=&(\omega_{ab}\omega^{AB}Z_{1B}^bZ_{2A}^a)_{{\cal N}=5}\\
\nonumber &\rightarrow&\omega_{a\alpha, b\beta}
\omega^{AB}Z_{1B}^{b\beta}Z_{2A}^{a\alpha}\\
&=&\bar Z_{2A}^aZ^A_{1a}+\bar Z_{1A}^aZ^A_{2a}.
\end{eqnarray}
Namely, it becomes a sum of the hermitian bilinear form of two
${\cal N}=6$ fields and its complex conjugate. Generally speaking,
the hermitian bilinear form of two arbitrary ${\cal N}=6$ 3-algebra
valued fields will become
\begin{eqnarray}\label{N6InnerProd}
h(X,Y)=X^*_aY_a\equiv \bar{X}^aY_a.
\end{eqnarray}

The reality condition (\ref{RealCondiOnN6F}) and Eq.
(\ref{DecomStrucConst}) imply that the ${\cal N}=5$ 3-bracket
(\ref{Symp3Bracket}) can be decomposed as a direct sum of ${\cal
N}=6$ brackets and their complex conjugates as follows:
\begin{eqnarray}\label{DecomBracket}\nonumber
[T_a,T_b; T_c]_{{\cal N}=5}&\rightarrow &[T_{a\alpha},T_{b\beta};
T_{c\gamma}]\\
\nonumber&=&[t^a,t^c;\bar{t}_b]\delta_{2\alpha}\delta_{1\beta}\delta_{2\gamma}
+[t^a,t^c;\bar{t}_b]^*\delta_{1\alpha}\delta_{2\beta}\delta_{1\gamma}\\
&&
+[t^b,t^c;\bar{t}_a]\delta_{1\alpha}\delta_{2\beta}\delta_{2\gamma}
+[t^b,t^c;\bar{t}_a]^*\delta_{2\alpha}\delta_{1\beta}\delta_{1\gamma}.
\end{eqnarray}
Here the 3-brackets
\begin{eqnarray}\label{N6Bracket}
[t^a,t^c;\bar{t}_b]=f^{ac}{}_{bd}t^d .
\end{eqnarray}
are those for the ${\cal N}=6$ 3-algebra. Such 3-brackets were first
proposed by Bagger and Lambert \cite{Bagger08:3Alg} for a ${\cal
N}=6$ CSM theory. An unusual feature of the 3-brackets is that it
involves complex conjugate for the third generator. Our above
decomposition from the ${\cal N}=5$  3-algebra reveals clearly the
origin of the need for complex conjugation of the third generator.

Later we will see that the structure constants defined in Eq.
(\ref{N6Bracket}) are indeed \emph{anti-symmetric} in the first two
indices. (See Eq. (\ref{SymmeOfN6F}).) With Eq.
(\ref{DecomBracket}), the fundamental identity (\ref{FFI}) reduces
to
\begin{eqnarray}\label{N6FI}
f^{fc}{}_{dg}f^{ag}{}_{eb}-f^{af}{}_{gb}f^{gc}{}_{de}
+f^{cf}{}_{eg}f^{ag}{}_{db}-f^{ac}{}_{gb}f^{gf}{}_{ed}=0 ,
\end{eqnarray}
as desired. Also the constraint condition
(\ref{ConstraintOn3Bracket}) on the structure constants and the
symmetry properties (\ref{SymmeOfF}) of the structure constants
reduce to
\begin{eqnarray}\label{SymmeOfN6F}
f^{ab}{}_{cd}=-f^{ba}{}_{cd}=f^{ba}{}_{dc} .
\end{eqnarray}
One easily recognizes that eqs. (\ref{N6InnerProd}),
(\ref{N6Bracket}), (\ref{N6FI}), (\ref{RealCondiOnN6F}), and
(\ref{SymmeOfN6F}) are those defining the ${\cal N}=6$ 3-algebra
used in ref. \cite{Bagger08:3Alg}. (The relation between the ${\cal
N}=6$ 3-algebra and super Lie algebra was discussed in Ref.
\cite{Jakob}.)

Substituting Eq. (\ref{DecomScaFld}), (\ref{DecomFerFld}),
(\ref{DecomStrucConst}), and (\ref{DecomGauFld}) into the ${\cal
N}=5$ Lagrangian (\ref{GeneN5Lagran}) and the SUSY transformation
law (\ref{GeneSusyTransLaw}), and using the $Sp(4)$ identity
(\ref{Sp4Id1}) and (\ref{Sp4Id2}), we reproduces the ${\cal N}=6$
Lagrangian
\begin{eqnarray}\label{N6Lagrangian}
\nonumber {\cal L} &=& -D_\mu
\bar{Z}_A^aD^\mu Z^A_a - i\bar\psi^{Aa}\gamma^\mu D_\mu\psi_{Aa}\\
\nonumber && -if^{ab}{}_{cd}\bar\psi^{Ad} \psi_{Aa}
Z^B_b\bar{Z}_B^c+2if^{ab}{}_{cd}\bar\psi^{Ad}
\psi_{Ba}Z^B_b\bar{Z}_A^c\\ \nonumber
&&-\frac{i}{2}\varepsilon_{ABCD}f^{ab}{}_{cd}\bar\psi^{Ac}
\psi^{Bd}Z^C_aZ^D_b -\frac{i}{2}\varepsilon^{ABCD}f^{cd}{}_{ab}
\bar\psi_{Ac}\psi_{Bd}\bar{Z}_C^a\bar{Z}_D^b \\
&&+\frac{1}{2}\varepsilon^{\mu\nu\lambda}
(f^{ab}{}_{cd}A_{\mu}{}^c{}_b\partial_\nu A_{\lambda}{}^d{}_a
+\frac{2}{3}f^{ac}{}_{dg}f^{ge}{}_{fb}
A_{\mu}{}^b{}_aA_{\nu}{}^d{}_c A_{\lambda}{}^f{}_e)\\ \nonumber &&
-\frac{2}{3}(f^{ab}{}_{cd}f^{ed}{}_{fg}
-\frac{1}{2}f^{eb}{}_{cd}f^{ad}{}_{fg})\bar{Z}_A^c Z^A_e\bar{Z}_B^f
Z^B_a\bar{Z}_D^g Z^D_b ,
\end{eqnarray}
and the  ${\cal N}=6$ SUSY transformation law reads
\begin{eqnarray}\label{N6susy}
\nonumber  \delta Z^A_d &=& -i\bar\epsilon^{AB}\psi_{Bd} \\
 \nonumber
 \delta \bar{Z}_{A}^{d}
&=& -i\bar\epsilon_{AB}\psi^{Bd} \\
 \nonumber
\delta \psi_{Bd} &=& \gamma^\mu D_\mu Z^A_d\epsilon_{AB} +
  f^{ab}{}_{cd}Z^C_aZ^A_b \bar{Z}_{C}^{c} \epsilon_{AB}+f^{ab}{}_{cd}
  Z^C_a Z^D_{b} \bar{Z}_{B}^{c}\epsilon_{CD} \\
\delta \psi^{Bd} &=& \gamma^\mu D_\mu \bar{Z}_A^d\epsilon^{AB} +
  f^{cd}{}_{ab}\bar{Z}_C^a \bar{Z}_A^b Z^{C}_{c} \epsilon^{AB}
+f^{cd}{}_{ab}\bar{Z}_C^a \bar{Z}_D^{b} Z^{B}_{c}\epsilon^{CD} \\
\nonumber
 \delta \tilde A_\mu{}^c{}_d &=&
-i\bar\epsilon_{AB}\gamma_\mu Z^A_a\psi^{Bb} f^{ca}{}_{bd} +
i\bar\epsilon^{AB}\gamma_\mu \bar{Z}_{A}^{a}\psi_{Bb}f^{cb}{}_{ad}.
\end{eqnarray}
Here the SUSY transformation parameters $\epsilon_{AB}$ satisfy
\begin{eqnarray}
\epsilon_{AB}&=&-\epsilon_{BA}\\
\epsilon^*_{AB}&=&\epsilon^{AB}
=\frac{1}{2}\varepsilon^{ABCD}\epsilon_{CD}
\end{eqnarray}

Now the parameters $\epsilon_{AB}$ transform as the $\bf 6$ of
$SU(4)$. It is in this sense that the global ${\cal N}=5$ SUSY gets
enhanced to ${\cal N}=6$. The Lagrangian (\ref{N6Lagrangian}) and
the transformation law (\ref{N6susy}) are the same as the ones
obtained in the 3-algebra approach for ${\cal N}=6$ theories in ref.
\cite{Bagger08:3Alg}.

The ${\cal N}=6$ superconformal CSM theories in three dimensions can
be classified by super Lie algebras \cite{GaWi, Hosomichi:2008jb,
Kac} or by using group theory \cite{Schn}. Two primary types are
allowed: with gauge group $U(M)\times U(N)$ and $Sp(2N)\times U(1)$,
respectively. In the next two subsections we will drive these two
theories by specifying the structure constants of the ${\cal N}=6$
3-algebra.

\subsection{${\cal N}=6$, $Sp(2N)\times U(1)$}\label{Sp2NU1}

The Lagrangian and SUSY transformation law for this theory were
first constructed in ref. \cite{ChenWu1}, starting from a formalism
for the symplectic 3-algebra, involving an anti-symmetric tensor,
that is different from the 3-algebra formalism of Bagger and Lambert
\cite{Bagger08:3Alg}. Here we would like to present the theory
completely in the framework of ref. \cite{Bagger08:3Alg}. We first
specify the structure constants as \footnote{In the Lagrangian
(\ref{N6Lagrangian}) of section \ref{N6}, the index $a$ runs from 1
to $L$. In this subsection, we split it into two indices:
$a\rightarrow a\pm$, and set $L=4N$. We hope this will not cause any
confusion.}
\begin{equation}\label{SyplStruc1}
\omega_{a-,e+}\omega_{b-,f+}f^{e+,f+}{}_{c+,d+} =
-k[(\omega_{ab}\omega_{cd}
+\omega_{ac}\omega_{bd})h_{+-}h_{+-}+(\omega_{ad}\epsilon_{+-})(\omega_{bc}
\epsilon_{+-})],
\end{equation}
where $k$ is a real constant, $\omega^{ab}$ an antisymmetric
bilinear form ($a,b=1,2,\cdots,2N$), $h_{+-}=h_{-+}=1$ and
$\epsilon_{+-}=-\epsilon_{-+}=ih_{+-}$. Here $a,b$ are the $Sp(2N)$
indices while $+,-$ the $SO(2)$ indices. And $\omega_{a-,e+}\equiv
\omega_{ae}h_{-+}$ is the gauge invariant antisymmetric tensor.
Since $\omega_{a-,e+}$ is non-singular, Eq. (\ref{SyplStruc1}) is
equivalent to the following equation:
\begin{eqnarray}
f^{a+,b+}{}_{c+,d+} &=&k(\omega^{ab}\omega_{cd} +\delta^a{}_d\delta^b{}_c
-\delta^a{}_c{}\delta^b{}_d)\delta^{+}{}_{+}\delta^{+}{}_{+}.
\end{eqnarray}
Suppressing the $SO(2)$ indices gives
\begin{eqnarray}\label{SyplStruc}
f^{ab}{}_{cd} &=&k(\omega^{ab}\omega_{cd} +\delta^a{}_d\delta^b{}_c
-\delta^a{}_c{}\delta^b{}_d).
\end{eqnarray}
It is not too difficult to check that the structure constants
satisfy the FI (\ref{N6FI}) and the reality condition
(\ref{RealCondiOnN6F}), and also have the desired symmetry
properties (\ref{SymmeOfN6F}). We see that \emph{after} suppressing
the $SO(2)$ indices, the structure constants are the same as the
components of an embedding tensor in Ref. \cite{Bergshoeff}.

In fact, in accordance with Eqs. (\ref{N6PhyGaugFld}) and
(\ref{SyplStruc}), the gauge fields can be decompoesed into two
parts:
\begin{eqnarray}
\tilde A_\mu{}^c{}_d &=& A_\mu{}^b{}_af^{ca}{}_{bd}\\
\nonumber
&=&-(A_{\mu d}{}^{c}+A_\mu{}^c{}_d)+(A_\mu{}^a{}_a)\delta^c{}_d \\
\nonumber &\equiv & B_\mu{}^c{}_d+A_\mu\delta^c{}_d.
\end{eqnarray}
It is natural to identify $A_\mu$ as the $U(1)$ part of the gauge
potential, and $B_\mu{}^c{}_d$ as the $Sp(2N)$ part. The reason is
that we can recast $B_\mu{}^c{}_d$ as $A_\mu^{ab}(t_{ab})^c{}_d$,
where $(t_{ab})^c{}_d$ is in the fundamental representation of the
Lie algebra of $Sp(2N)$.

We substitute the structure constants (\ref{SyplStruc}) into (\ref{N6susy}). We then obtain the ${\cal N}=6$ (on-shell) SUSY transformation law in
the theory (see Appendix B.1). The equations of motion can be
derived from the Lagrangian obtained by substituting eqs.
(\ref{SyplStruc}) into the Lagrangian (\ref{N6Lagrangian}) and
replacing $ A_\mu{}^b{}_a$ by $\frac{1}{k} A_\mu{}^b{}_a$ (see
Appendix B.1). The SUSY transformation law (\ref{SyplSusyTrasf}) and
the Lagrangian (\ref{SymplLagrangian}) are indeed in agreement with
the ${\cal N}=6, Sp(2M)\times U(1)$ superconformal CSM theory
derived from the symplectic 3-algebra in ref. \cite{ChenWu1}, or
from the ordinary Lie algebra in ref. \cite{Hosomichi:2008jb}.

\subsection{${\cal N}=6$, $U(M)\times U(N)$}
\label{UMUN}

The Lagrangian this theory has been constructed in ref.
\cite{Bagger08:3Alg}. For this paper to be self-contained, it is
worth to present the Lagrangian and SUSY transformation law of $D=3,
{\cal N}=6$, $U(M)\times U(N)$ theory in this subsection.

To generate a direct gauge group such as $U(M)\times U(N)$, we split
up a \emph{lower} 3-algebra index $a$ into two indices:
$a\rightarrow n\hat n$, where $n=1,...,M$ is a fundamental index of
$U(M)$, $\hat n= 1,...,N$ an anti-fundamental index of $U(N)$. With
this decomposition, the hermitian inner product (\ref{N6InnerProd})
can be written as a trace:
\begin{equation}
X^*_aY_{a}\rightarrow X^*_{n\hat{n}}Y_{n\hat{n}}= X^{*{\rm
t}}_{\hat{n}n}Y_{n\hat{n}}\equiv {\rm Tr}(X^\dag Y) ,
\end{equation}
where the superscript ``t" stands for the usual transpose. On the
other hand, according to the definition (\ref{N6InnerProd}), the
hermitian inner product can be also written as: $X^*_aY_{a}\equiv
\bar{X}^aY_{a}$, which leads us to decompose an \emph{upper} index
$a$ as $a\rightarrow \hat nn$. Thus the hermitian inner product can
be written as
\begin{equation}
X^*_aY_{a}\equiv \bar{X}^aY_{a}\rightarrow \bar{X}^{\hat nn}Y_{n\hat
n}\equiv {\rm Tr}(\bar{X}Y)={\rm Tr}(X^\dag Y).
\end{equation}
We then specify the 3-bracket $(\ref{N6Bracket})$ to be
\begin{eqnarray}\label{Unitary3Bracket}
[t^{\hat kk}, t^{\hat ll}; \bar{t}_{m\hat{m}}]=k(\delta^{\hat
k}{}_{\hat m}\delta^l{}_m t^{\hat lk}-\delta^{\hat l}{}_{\hat
m}\delta^k{}_m t^{\hat kl}) .
\end{eqnarray}
The structure constants can be easily read off as
\begin{eqnarray}\label{UnitaryStruc}
f^{\hat kk,\hat ll}{}_{m\hat{m},n\hat n}=k(\delta^{\hat k}{}_{\hat
m}\delta^{\hat l}{}_{\hat n}\delta^k{}_n\delta^l{}_m -\delta^{\hat
k}{}_{\hat n}\delta^{\hat l}{}_{\hat m}\delta^k{}_m\delta^l{}_n).
\end{eqnarray}
It is straightforward to check that the structure constants $f^{\hat
kk,\hat ll}{}_{m\hat{m},n\hat n}$ satisfy the FI (\ref{N6FI}) and
the reality conditions (\ref{RealCondiOnN6F}), and has the symmetry
properties (\ref{SymmeOfN6F}). The structure constants are first
discovered by BL \cite{Bagger08:3Alg} (though they did not write
down Eq. (\ref{UnitaryStruc}) explicitly), and they are also the
same as the components of an embedding tensor in Ref.
\cite{Bergshoeff}.

Now let us show that the 3-bracket (\ref{Unitary3Bracket}) is indeed
equivalent to Bagger and Lambert's 3-bracket \cite{Bagger08:3Alg}.
Writing $X=X_{k\hat k}t^{\hat kk}$, and $\bar Z=\bar{Z}^{\hat
mm}\bar{t}_{m\hat{m}}$, by Eq. (\ref{Unitary3Bracket}), one can get
\begin{eqnarray}\label{BL3Bracket}
[X, Y; \bar{Z}]=k(X\bar{Z}Y-Y\bar{Z}X)_{n\hat n} t^{\hat nn} .
\end{eqnarray}
The right hand side is the ordinary matrix multiplication. It is
exactly the same as Eq. (53) of Ref. \cite{Bagger08:3Alg}. In
accordance with eq. (\ref{UnitaryStruc}), the gauge fields can be
decomposed as
\begin{eqnarray}
\nonumber\tilde{A}_\mu{}^{\hat kk}{}_{n\hat n}&=& A_\mu{}^{\hat
mm}{}_{l\hat l}f^{\hat kk,\hat ll}{}_{m\hat{m},n\hat n}\\
\nonumber &=&A_\mu{}^{\hat kl}{}_{l\hat n}\delta^k{}_n-A_\mu{}^{\hat
lk}{}_{n\hat l }\delta^{\hat k}{}_{\hat n}\\
&\equiv& \hat{A}_\mu{}^{\hat k}{}_{\hat n}\delta^k{}_n
+A_\mu{}^{k}{}_{n}\delta^{\hat k}{}_{\hat n} .
\end{eqnarray}
So the 3-bracket (\ref{BL3Bracket}) and the FI (\ref{N6FI}) generate
a $U(M)\times U(N)$ gauge group \cite{Bagger08:3Alg}, with
$\hat{A}_\mu{}^{\hat k}{}_{\hat n}$ the $U(M)$ part and
$A_\mu{}^{k}{}_{n}$ the $U(N)$ part of the gauge potential.

The supersymmetry transformation law and the Lagrangian in this
theory can be obtained by substituting the expression
(\ref{UnitaryStruc}) of the structure constants into eqs.
(\ref{N6susy}) and (\ref{N6Lagrangian}), and replacing
$A_\mu{}^b{}_a$ by $\frac{1}{k} A_\mu{}^b{}_a$. To make the paper
self-contained, we include the results in Appendix B.2. The SUSY
transformation law (\ref{UnitarySusyTrasf}) and the Lagrangian
(\ref{ABJM}) are in agreement with the $D=3, {\cal N}=6$ $U(M)\times
U(N)$ CSM theory, which has been derived from the ordinary Lie
algebra approach in ref. \cite{Hosomichi:2008jb} and from the
3-algebra approach in ref. \cite{Bagger08:3Alg}.

This theory is conjectured to be the dual gauge theory of M2 branes
a $\textbf{C}^4/\textbf{Z}_k$ singularity. If $M=N$, this theory
becomes the well-known ABJM model \cite{ABJM, Benna, Schwarz}.

\section{Conclusions}\label{Conclusions}
In this paper, we first introduce an anti-symmetric tensor
$\omega_{ab}$ into a 3-algebra, with structure constants of the
3-bracket being \emph{symmetric} in the first two indices. We call
it a symplectic 3-algebra.
We then construct the general ${\cal N}=5$ superconformal CSM theory
with $Sp(4)$ R-symmetry in three dimensions in terms of this
symplectic 3-algebra. All matter fields take values in this
symplectic 3-algebra. The gauge symmetry is generated by the
3-bracket and FI. By specifying the 3-brackets, we provide the
${\cal N}=5, Sp(2N)\times O(M)$ CSM theory as an example of our
theory. It would be nice to see if CSM theories with other gauge
groups (for example, $G_2\times SU(2)$ \cite{Bergshoeff}) for
multiple M2 branes can be generated in a similar way with the
3-algebra approach. Also it would be interesting to generalize the
symplectic 3-algebra theory, so that it can describe CSM quiver
gauge theories.

We have succeeded in enhancing the ${\cal N}=5$ supersymmetry to
${\cal N}=6$ by decomposing the sympelctic 3-algebra and the fields
properly. At the same time, we also demonstrate that the FI and the
symmetry and reality properties of the structure constants of the
${\cal N}=6$ 3-algebra can be derived from the ${\cal N}=5$
counterparts. Hence some of ${\cal N}=5, 6$ superconformal CSM
theories are described by a unified sympletic 3-algebraic framework.
It would be nice to investigate the relation between these two kinds
of 3-algebras further. By specifying the 3-brackets, the ${\cal
N}=6$, $Sp(2N)\times U(1)$ and $U(M)\times U(N)$ CSM, including the
ABJM theory, are derived. We also compare the approach used in our
previous paper \cite{ChenWu1} with that of this paper. The same
theory ($Sp(2N)\times U(1)$) are derived by starting from different
3-algebra formalisms.

It would be nice to construct the ${\cal N}\leq4$ superconformal CSM
theories \cite{MFM:Aug09, Saemann2} for multiple M2 branes in terms
of 3-algebras.

\section{Acknowledgement} We would like to thank Yong-Shi Wu for
useful discussions. We also thank the referee for comments.

\appendix

\section{Conventions and Useful Identities}\label{Identities}

In $1+2$ dimensions, the gamma matrices are defined as
$\{\gamma_\mu, \gamma_\nu\}= 2\eta_{\mu\nu}$. For the metric we use
the $(-,+,+)$ convention. The gamma matrices can be defined as the
Pauli matrices: $\gamma_\mu=(i\sigma_2, \sigma_1, \sigma_3)$,
satisfying the important identity
\begin{equation}
\gamma_\mu\gamma_\nu=\eta_{\mu\nu}+\varepsilon_{\mu\nu\lambda}\gamma^{\lambda}.
\end{equation}
We also define
$\varepsilon^{\mu\nu\lambda}=-\varepsilon_{\mu\nu\lambda}$. So
$\varepsilon_{\mu\nu\lambda}\varepsilon^{\rho\nu\lambda} =
-2\delta_\mu{}^\rho$. The Fierz transformation is
\begin{equation}
(\bar\lambda\chi)\psi = -\frac{1}{2}(\bar\lambda\psi)\chi
-\frac{1}{2} (\bar\lambda\gamma_\nu\psi)\gamma^\nu\chi.
\end{equation}
Some useful $Sp(4)$ identities are
\begin{eqnarray}
\bar\epsilon^{AC}_1\epsilon_{2C}{}^{B}
-\bar\epsilon^{AC}_2\epsilon_{1C}{}^{B}&=&
\bar\epsilon^{BC}_1\epsilon_{2C}{}^{A}
-\bar\epsilon^{BC}_2\epsilon_{1C}{}^{A}\\
\frac{1}{2}\bar\epsilon^{CD}_1\gamma_\nu\epsilon_{2CD}\,\delta^A_B
&=&\bar\epsilon^{AC}_1\gamma_\nu\epsilon_{2BC}
-\bar\epsilon^{AC}_2\gamma_\nu\epsilon_{1BC}\\
 \nonumber
2\bar\epsilon^{AC}_1\epsilon_{2BD}
-2\bar\epsilon^{AC}_2\epsilon_{1BD}
&=&\bar\epsilon^{CE}_1\epsilon_{2DE}\delta^A_B
-\bar\epsilon^{CE}_2\epsilon_{1DE}\delta^A_B\\
\nonumber
&-&\bar\epsilon^{AE}_1\epsilon_{2DE}\delta^C_B
+\bar\epsilon^{AE}_2\epsilon_{1DE}\delta^C_B\\
&+& \bar\epsilon^{AE}_1\epsilon_{2BE}\delta^C_D
-\bar\epsilon^{AE}_2\epsilon_{1BE}\delta^C_D\\
\nonumber
&-& \bar\epsilon^{CE}_1\epsilon_{2BE}\delta^A_D
+\bar\epsilon^{CE}_2\epsilon_{1BE}\delta^A_D\\
\nonumber \frac{1}{2}\varepsilon_{ABCD}
\bar\epsilon^{EF}_1\gamma_\mu\epsilon_{2EF}
&=&\bar\epsilon_{1AB}\gamma_\mu\epsilon_{2CD}
-\bar\epsilon_{2AB}\gamma_\mu\epsilon_{1CD}\\
&+& \bar\epsilon_{1AD}\gamma_\mu\epsilon_{2BC}
-\bar\epsilon_{2AD}\gamma_\mu\epsilon_{1BC}\\
\nonumber
&-&\bar\epsilon_{1BD}\gamma_\mu\epsilon_{2AC}
+\bar\epsilon_{2BD}\gamma_\mu\epsilon_{1AC}\\
\varepsilon^{ABCD}&=&-\omega^{AB}\omega^{CD}
+\omega^{AC}\omega^{BD}-\omega^{AD}\omega^{BC}.
\end{eqnarray}
The $Sp(4)$ indices can lowered and raised by the anti-symmetric
tensor $\omega_{AB}$ and its inverse $\omega^{AB}$.

\section{SUSY Transformation Law and Lagrangian in
$D=3$, ${\cal N}=6$ CSM Theories}

For this paper to be self contained, below we give the explicit form
of the SUSY transformation law and the Lagrangian for the $D=3,
{\cal N}=6$ CSM theories with $SU(4)$ $R$-symmetry. For the
notations, see the corresponding subsections in the text.

\subsection{$Sp(2N)\times U(1)$ CSM Theory}

The Lagrangian of the theory is given by
\begin{eqnarray}\label{SymplLagrangian}
\nonumber {\cal L} &=& -D_\mu \bar{Z}_A^aD^\mu Z^A_a -
i\bar\psi^{Aa}\gamma^\mu D_\mu\psi_{Aa} \\
&& +ik(\bar{Z}^b_B\bar \psi_{Ab}\psi^{Aa}Z^B_a-\bar Z_B^b
Z^B_b\bar\psi^{Aa}\psi_{Aa}-\bar
Z_B^c\omega_{cd}\bar\psi^{Ad}\psi_{Aa}\omega^{ab}Z^B_b)\\
\nonumber && -2ik(\bar{Z}^b_B\bar \psi_{Ab}\psi^{Ba}Z^A_a-\bar Z_B^b
Z^A_b\bar\psi^{Ba}\psi_{Aa}-\bar
Z_B^c\omega_{cd}\bar\psi^{Bd}\psi_{Aa}\omega^{ab}Z^A_b)\\
\nonumber && -ik\varepsilon^{ABCD}(\bar{Z}^a_A\bar \psi_{Ba}\bar
Z_C^b\psi_{Db}-\frac{1}{2}\bar Z_A^c\omega_{cd}\bar
Z_C^d\bar\psi_{Ba}\omega^{ab}\psi_{Db})\\
\nonumber &&-ik\varepsilon_{ABCD}(\bar\psi^{Ba}Z_a^A\psi^{Db}Z^C_b
-\frac{1}{2}Z^A_a\omega^{ab} Z^C_b\bar\psi^{Bc}
\omega_{cd}\psi^{Dd})\\ \nonumber
&&+\frac{1}{2k}\varepsilon^{\mu\nu\lambda}A_\mu\partial_\nu
A_\lambda-\frac{1}{4k}\varepsilon^{\mu\nu\lambda}
{\rm Tr}(B_\mu\partial_\nu B_\lambda +\frac{2}{3}B_\mu B_\nu B_\lambda)\\
\nonumber &&-3k^2Z^B_a\omega^{ab}Z^D_b\bar Z_D^eZ^A_e \bar
Z_A^c\omega_{cd}\bar Z_B^d
+\frac{5k^2}{3}\bar Z_A^aZ^B_a \bar Z_B^bZ^D_b\bar Z_D^cZ^A_c\\
\nonumber &&-2k^2\bar Z_A^aZ^B_a\bar Z_D^bZ^D_b\bar Z_B^c Z^A_c
+\frac{k^2}{3}\bar Z_B^aZ^B_a\bar Z_D^bZ^D_b\bar Z_A^c Z^A_c .
\end{eqnarray}

The SUSY transformation laws are given by
\begin{eqnarray}\label{SyplSusyTrasf}
\nonumber  \delta Z^A_d &=& -i\bar\epsilon^{AB}\psi_{Bd} \\
 \nonumber
 \delta \bar{Z}_{A}^{d} &=& -i\bar\epsilon_{AB}\psi^{Bd} \\
 \nonumber
\delta \psi_{Bd} &=& \gamma^\mu D_\mu Z^A_d\epsilon_{AB} -
  kZ^C_a\omega^{ab}Z^A_b \omega_{dc}\bar{Z}_{C}^{c} \epsilon_{AB}-k
  Z^C_a \omega^{ab}Z^D_{b}\omega_{dc}\bar{Z}_{B}^{c}\epsilon_{CD}
\\ \nonumber
  &&-k\bar Z_C^a Z^C_a Z^A_d\epsilon_{AB}+k\bar Z_C^a Z^A_a
  Z^C_d\epsilon_{AB}-2k\bar Z_B^a Z^C_a Z^D_d\epsilon_{CD}
\\ \nonumber
\delta \psi^{Bd} &=& \gamma^\mu D_\mu \bar{Z}_A^d\epsilon^{AB} -
k\bar{Z}_C^a \omega_{ab}\bar{Z}_A^b \omega^{dc}Z^{C}_{c}
\epsilon^{AB} -k\bar{Z}_C^a \omega_{ab}\bar{Z}_D^{b}
\omega^{dc}Z^{B}_{c}\epsilon^{CD}
\\ \nonumber &&-k\bar Z_C^a Z^C_a \bar Z_A^d\epsilon^{AB}
+k\bar Z_A^a Z^C_a \bar Z_C^d\epsilon^{AB} -2k\bar Z_C^a Z^B_a \bar
Z_D^d\epsilon^{CD}\\ \nonumber
 \delta A_\mu &=&
-ik\bar\epsilon_{AB}\gamma_\mu \psi^{Ba}Z^A_a+
ik\bar\epsilon^{AB}\gamma_\mu \bar{Z}_{A}^{a}\psi_{Ba}\\
\nonumber \delta B_\mu{}^{c}{}_d &=& ik\bar\epsilon_{AB}\gamma_\mu
\omega^{ca}Z^A_a\omega_{db}\psi^{Bb} -i k\bar\epsilon^{AB}\gamma_\mu
\omega_{db}\bar Z_A^b\omega^{ca}\psi_{Ba}\\
&& +ik\bar\epsilon_{AB}\gamma_\mu Z^{A}_{d}\psi^{Bc}
-ik\bar\epsilon^{AB}\gamma_\mu \bar Z_{A}^{c}\psi_{Bd}.
\end{eqnarray}

\subsection{$U(M)\times U(N)$ CSM Theory}

The Lagrangian of the theory is given by
\begin{eqnarray}\label{ABJM}
\nonumber {\cal L} &=& -{\rm Tr}(D_\mu \bar{Z}_A D^\mu Z^A) -
i{\rm Tr}(\bar\psi^{A}\gamma^\mu D_\mu\psi_A) -V+{\cal L}_{CS}\\
\nonumber && -ik{\rm Tr}(\bar\psi^{A} \psi_{A} \bar{Z}_B
Z^B-\bar\psi^{A} Z^B \bar{Z}_B\psi_{A})\\
 &&+2ik {\rm Tr}(\bar\psi^{A}\psi_{B} \bar{Z}_A Z^B
 -\bar\psi^{A} Z^B \bar{Z}_A\psi_{B})\\ \nonumber
 &&+ik\varepsilon_{ABCD}{\rm Tr}(\bar\psi^{A} Z^C\bar\psi^{B}Z^D)
 -ik\varepsilon^{ABCD}{\rm Tr}(\bar{Z}_D \psi_A \bar{Z}_C\psi_B)\ .
\end{eqnarray}
The Lagrangian (\ref{ABJM}) is the same obtained by BL
\cite{Bagger08:3Alg}, except for that we re-scale the gauge fields by
a factor $\frac{1}{k}$. The potential term is
\begin{eqnarray}
\nonumber V&=&2k^2{\rm Tr}(\bar Z_AZ^A\bar Z_BZ^C\bar
Z_CZ^B)-\frac{4k^2}{3}{\rm Tr}(Z^A\bar Z_BZ^C\bar Z_AZ^B\bar Z_C)\\
&&-\frac{k^2}{3}{\rm Tr}(Z^A\bar Z_AZ^B\bar Z_BZ^C\bar Z_C+\bar
Z_AZ^A\bar Z_BZ^B\bar Z_CZ^C).
\end{eqnarray}
The Chern-Simons term reads
\begin{eqnarray}
{\cal L}_{CS}=\frac{1}{2k}\varepsilon^{\mu\nu\lambda}{\rm
Tr}\bigg(\hat{A}_ \mu\partial_\nu\hat{A}_\lambda+\frac{2}{3}\hat{A}_
\mu\hat{A}_\nu\hat{A}_\lambda-A_ \mu\partial_\nu
A_\lambda-\frac{2}{3}A_ \mu A_\nu A_\lambda\bigg) .
\end{eqnarray}

The ${\cal N}=6$ SUSY transformation laws, which are closed on-shell
with the equations of motion derivable from the above lagrangian
(\ref{ABJM}), are given by
\begin{eqnarray}\label{UnitarySusyTrasf}
\nonumber  \delta Z^A &=& -i\bar\epsilon^{AB}\psi_{B} \\
 \nonumber
 \delta \bar{Z}_{
A}&=& -i\bar\epsilon_{AB}\bar{\psi}^{B} \\
\nonumber \delta \psi_{B} &=& \gamma^\mu D_\mu Z^A\epsilon_{AB} +
k(Z^C\bar{Z}_C Z^A-Z^A\bar{Z}_C
Z^C)\epsilon_{AB}+2kZ^C\bar{Z}_BZ^D\epsilon_{CD}\\ \nonumber \delta
\bar{\psi}^{B} &=& \gamma^\mu D_\mu \bar{Z}_A\epsilon^{AB}+
k(\bar{Z}_AZ^C\bar{Z}_C -\bar{Z}_CZ^C\bar{Z}_A)\epsilon^{AB}+2k\bar
{Z}_DZ^B\bar{Z}_C\epsilon^{CD}\\ \nonumber
\delta\hat{A}_\mu&=&-ik\bar{\epsilon}_{AB}\gamma_\mu\bar{\psi}^B
Z^A+ik\bar{\epsilon}^{AB}\gamma_\mu\bar{Z}_A\psi_B\\
\delta A_\mu&=&ik\bar{\epsilon}_{AB}\gamma_\mu Z^A\bar{\psi}^B
-ik\bar{\epsilon}^{AB}\gamma_\mu\psi_B\bar{Z}_A .
\end{eqnarray}

\end{document}